\documentclass[pmlr,twocolumn]{jmlr} 
\usepackage{booktabs}
\usepackage[load-configurations=version-1]{siunitx} 
\usepackage[pscoord]{eso-pic}
\pdfoutput=1

\newcommand{\placetextbox}[3]{
  \setbox0=\hbox{#3}
  \AddToShipoutPictureFG*{
    \put(\LenToUnit{#1\paperwidth},\LenToUnit{#2\paperheight}){\vtop{{\null}\makebox[0pt][c]{#3}}}
  }
}

\jmlrvolume{ML4H Extended Abstract Arxiv Index}
\jmlryear{2020}
\jmlrsubmitted{2020}
\jmlrpublished{}
\jmlrworkshop{Machine Learning for Health (ML4H) 2020}

\title[Recommendations for hierarchical model specifications for case-control studies]{\vspace{-5ex}Recommendations for Bayesian hierarchical model specifications for case-control studies in mental health}

\author{%
\Name{Vincent Valton} \Email{v.valton@ucl.ac.uk}\\
\addr Institute of Cognitive Neuroscience, University College London
\AND
\Name{Toby Wise} \Email{t.wise@ucl.ac.uk}\\
\addr Max Planck UCL Centre for Computational Psychiatry and Ageing Research,\\ University College London
\AND
\Name{Oliver J. Robinson} \Email{o.robinson@ucl.ac.uk}\\
\addr Institute of Cognitive Neuroscience, University College London\vspace{-5ex}
}

\begin{document}
\maketitle

\placetextbox{0.433}{0.08}{\scriptsize ML4H Workshop, 34th Conference on Neural Information Processing Systems (NeurIPS 2020)}

One in four people worldwide will experience a mental health disorder in their lifetime, and depressive disorders alone rank second in the leading causes of global disease burden worldwide \citep{WHO:2013}.  Mental health is one area of healthcare that is in dire need of improvement, and where the successful application of machine learning has the potential to reach and improve the lives of millions worldwide.  Of particular interest is hierarchical model fitting, which has become commonplace for case-control studies of cognition and behaviour in mental health \citep{Daw:2009p7819,Ahn:2017kp,Huys:2016jwa}. These methods allow practitioners to formalise differences in symptoms between patients and controls in terms of latent variables (parameters) from computational models \citep{Anonymous:26lP5Bz9}. By formalising the computations by which pathological symptoms emerge, researchers and practitioners are able to better understand the underlying mechanisms leading to these symptoms; better understand the role of current treatments in modulating these mechanisms; and hence improve our ability to target and develop new treatments.

These techniques, however, require us to formalise assumptions about the data-generating process at the group level, which may not be known. Two schools of thought currently exist: (1) - assume all subjects are drawn from a common population but at the risk of underestimating true group differences between patients and controls, leading to an increased rate of false negative findings; (2) - model both groups as deriving from separate populations at the risk of overestimating true group differences between patients and controls, leading to an increased rate of false positive findings. Neither of these assumptions has been formally evaluated and quantified, yet both inflated false negative or false positive findings can have disastrous implications: either missing potentially impactful avenues of research for patients, or conversely allocating scarce research resources and funding towards an impasse. It is therefore critical that we formally and carefully explore, analyse and quantify these different approaches to provide adequate recommendations for future mental health research.

\begin{figure*}[htbp]
\floatconts
  {fig:workflow}
  {\caption{Study workflow}}
  {\includegraphics[width=1\textwidth,height=8.2cm]{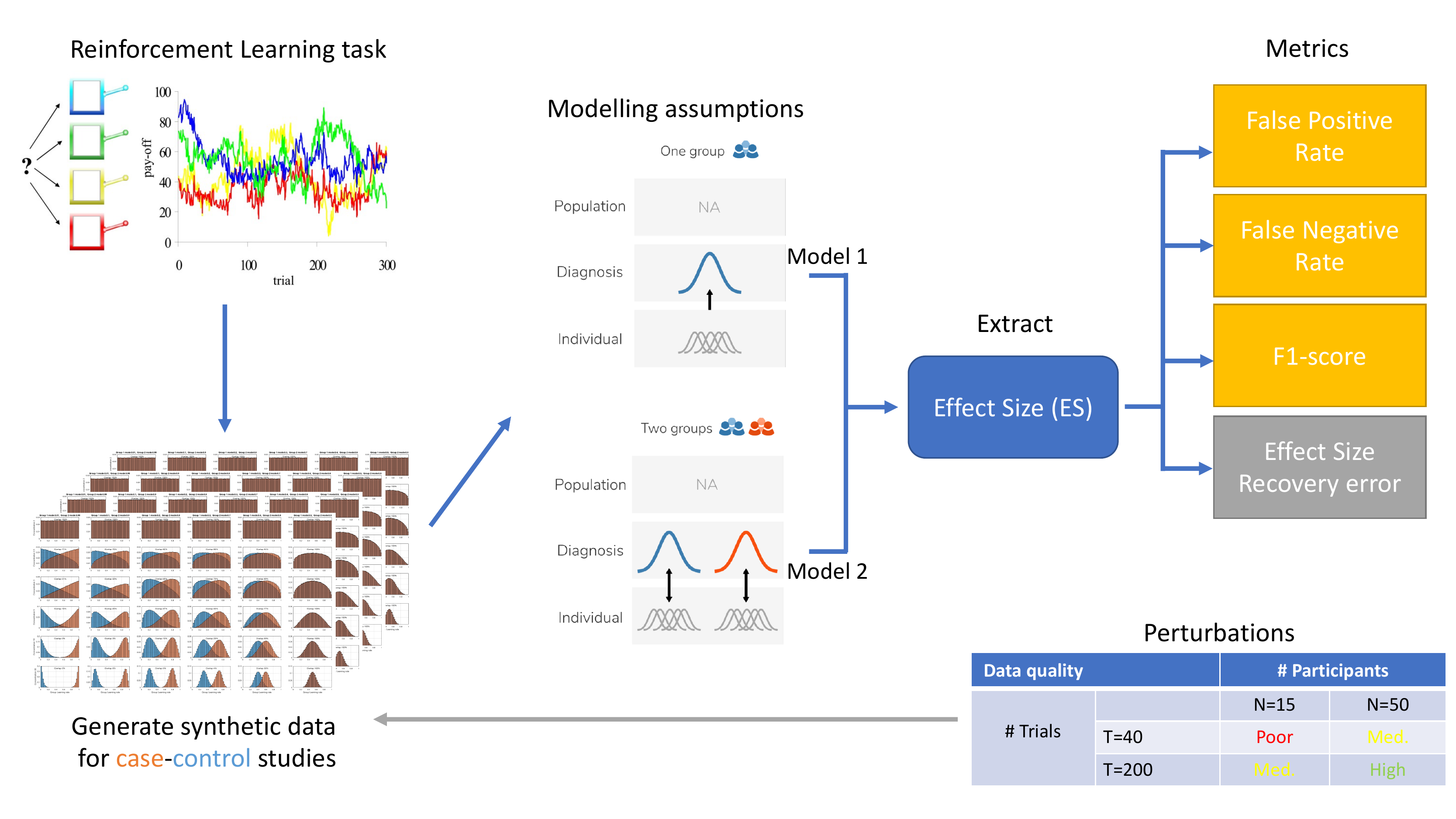}}
\end{figure*}

To address these concerns, we systematically tested these assumptions using simulations on synthetic data from a commonly used multi-armed bandit task (RL task - \citet{Seymour:2012cza}). We examined recovery of group differences in latent parameter space under different generative modelling assumptions: (\emph{Model 1}) modelling groups under a common shared group-level prior (assuming all participants are generated from a common distribution, and are likely to share common characteristics); (\emph{Model 2}) modelling separate groups based on symptomatology or diagnostic labels, resulting in separate group-level priors. We then evaluated the robustness of these approaches to perturbations in data quality.

We show that fitting groups separately (\emph{Model 2}) provides the most accurate recovery of true group differences between case and controls (F1 Score). Finally, we demonstrate that \emph{model 2} provides the most robust and unbiased recovery of effect sizes across all combinations of data perturbations (as measured by the absolute error in effect-size recovery). 

\section{Methods}
 
\paragraph{Task:} We opted to use a reinforcement learning (RL) task as our behavioural task (multi-armed bandit), as it is one of the most used and studied paradigm in computational psychiatry \citep{Anonymous:26lP5Bz9,Seymour:2012cza,Mkrtchian:2017hj, Aylward:2019ka, Wilson:2019kt, Brown:2020ie}. We chose to use roving probabilities as this allowed us to estimate latent parameters more accurately than conventional stationary RL environments. Specifically, stationary RL environments (e.g. with fixed 80/20\% reward probabilities) result in fewer usable trials to estimate learning rates, as participants will tend to stay with the best option once it has been identified. Using a multi-arm bandit with roving probabilities allows us to ensure maximal parameter recovery and to avoid introducing biases in the parameter recovery due to asymptotic performance at the task.

\paragraph{Synthetic Datasets:} We generated data using a typical RL model \citep{Dayan:2008p8403,Daw:2009p7819} for two separate groups (case \& controls), with varying degrees of overlap in parameter space across groups. To study the effect of different model specifications on model fit and parameter recovery, we opted to start with a sufficiently large number of subjects per group (n=50) and number of trials (t=200). This ensured that differences between recovery of model parameters were due to model specifications alone, and not due to insufficient data for parameter recovery. 
Beta distributions were used to ensure that the learning rate was constrained between zero and one. We then generated 36 synthetic datasets by modifying the two groups modes between the ranges of [0.01, 0.1:0.1:0.9, 0.99], and with concentration parameters varying in the range of [2, 2.5, 3, 5, 10, 30]. This range allowed us to simulate groups with complete overlap in parameter space, with varying degrees of similarity within each group (see figure \ref{fig:workflow}, bottom left corner). Using these datasets, we could then compare models of different specifications, and study their failure points under different hierarchical assumptions and different effect sizes. We then resampled each dateset 1000 times (total of 36,000 datasets) to allow for precise computations of model accuracy metrics.
For every synthetic dataset generated, we ensured that the resampling of parameters was within 0.02 absolute error of the theoretical data-generating distribution. This ensured that any bias observed during the recovery of group-level parameters (i.e. group mean and standard deviation) were due to model specifications, and not due to random unrepresentative samples from the data generating distribution. 

\paragraph{Data perturbations:} Finally, in order to assess the robustness of each model specification to data quality, we varied the number of subjects and trials in each dataset \citep{Baker:2020di}. Due to combinatorial issues, we opted to test only two group sizes, and two sets of trial sizes, which represent the limits of acceptable data quality. We assumed 50 subjects per group with 200 trials per subject as our optimal (High) data quality, and allowed for a lower group size of merely 15 subjects per group (as it is not uncommon to see such sample sizes in costly experimental settings, such as in fMRI studies \textemdash \citet{Nord:2017bn}). We opted for 40 trials as our lowest use-case for low data quality (as it is not unusual to see RL studies with no more than 2-4 pairs of stimuli with only 10-15 trials each in a stationary environment). We selected the sequence of 40 consecutive trials which resulted in the highest recovery of parameters (out of the original 200 trials), so that the analysis provided an upper bound as to what each model was able to achieve when using sub-optimal data. 

\paragraph{Modelling:} All models were implemented and fitted using the probabilistic language \emph{Stan} \citep{Gelman:2013una}, with 2 chains of 3000 samples per fit, including 1000 samples per chain used for warm up.
Group$-$level priors were calibrated across all models \citep{Gabry:2017go}, to ensure that any observed difference between models were due to the model specification alone. All models were rigorously tested for quality checks on uniformly distributed synthetic data to ensure that they could adequately recover parameters (Pearson $\rho > 0.92$). 

\begin{table*}[htbp]
\vspace{-6ex}
\floatconts{tab:F1}
{\caption{Model accuracy: False positive rate, false negative rate, F1-Score, and 95\% CI}\vspace{-2ex}}
{\begin{tabular}{lccr}
\toprule 
 {$50$ subj. $200$ trials} & False Pos. Rate (\%) & False Neg. Rate (\%) & F1-Score (\%)
 \tabularnewline
\midrule 
Model 1 & 0.48 [$\pm$ 0.07] & 6.03 [$\pm$ 0.24] & 96.73 [$\pm$ 0.23] \tabularnewline
Model 2 & 2.66 [$\pm$ 0.16] & \textbf{\textcolor{black}{1.75 [}} $\pm$ \textbf{\textcolor{black}{0.13]}} & \textbf{\textcolor{black}{98.26 [}}$\pm$ \textbf{\textcolor{black}{0.17]}} \tabularnewline
\bottomrule 
\end{tabular}
}
\end{table*}

\begin{table*}[htbp]
\floatconts{tab:ES_error}
{\caption{E.S. error (\%) - positive and negative sign denote over/under estimation}\vspace{-2ex}}
{\begin{tabular}{lcccc}
\toprule 
Data  & \multicolumn{4}{c}{Effect size recovery error (\%)}\tabularnewline
perturb. & \textit{\textcolor{black}{50 subj. 200 trials}} & \textit{\textcolor{black}{15 subj. 200 trials}} & \textit{\textcolor{black}{50 subj., 40 trials}} & \textit{\textcolor{black}{15 subj. 40 trials}}\tabularnewline
\midrule 
Model 1 & -28.60 & -17.74 & -63.81 & -64.46\tabularnewline
Model 2 & \textbf{\textcolor{black}{+7.85}} & \textbf{\textcolor{black}{+9.41}} & \textbf{\textcolor{black}{+28.12}} & \textbf{\textcolor{black}{+29.09}}\tabularnewline
\bottomrule 
\end{tabular}}
\end{table*}

\paragraph{Metrics:} One of the most important metric we want to recover from our computational analysis is the true parameter difference between case \& control groups. To do so, we need to accurately recover the group-level summary statistics so that we can adequately compare groups' means. Ideally, this would be done using fully Bayesian approaches \citep{Kruschke:2014wl,Ahn:2017kp}, but in practice this is often pragmatically done using standard general linear models (e.g. t-tests, anovas, etc.). 
One way to summarise parameter recovery for both groups using a single metric is to use the effect size of the difference between groups, as it naturally combines both group means and standard deviation, which themselves rely on the correct estimation of the subject-level parameters within each group. To compare each model specification, we therefore calculate the true effect size (Cohen’s d) between the two groups from the synthetic datasets, and then calculate the recovered effect-size between groups using each model specification. We also report important metrics of interest such as the false positive rate, the false negative rate, and the F1 score, as they allow us to summarise the accuracy and pitfalls for each model specification.

\section{Results \& Discussion}

\paragraph{Model accuracy:} We find that \emph{model 2} (fitting groups separately) provided better overall recovery of true group differences in parameter space than \emph{model 1} (fitting groups under the same prior), as summarized by the F1 score (see table \ref{tab:F1}). While \emph{model 2} resulted in a slightly increased rate of false positive findings, it led to a much lower rate of false negative findings than \emph{model 1}. This is because \emph{model 1} `pools' participants estimates towards the shared group-level prior (regularisation), leading to an underestimation of the effect size. 

\paragraph{Effect size recovery:} We were next interested in determining how data quality impacted parameter recovery, as it could be that some model specifications are particularly prone to error when data quality deteriorates. We focused here on the number of trials and the number of subjects, two factors that are often limited in psychiatric research.
\emph{Model 1} was particularly inadequate in the poor data-quality cases (low number of trials), in which it underestimated the true effect size by as much as -64\% (Table \ref{tab:ES_error}). In more optimal data quality settings \emph{model 1} still underestimated the true effect size, but by a more acceptable margin.
Conversely, \emph{model 2} allowed ‘pooling’ solely for the individuals belonging to the group being estimated. As a result, it was more accurate at recovering the effect size between the two groups (in absolute error terms). Data quality appeared to degrade the recovery of the effect size for both models, particularly when trial number decreased. While \emph{model 2} overestimated the effect size in high data quality settings (high number of trials), it was not as catastrophically affected as \emph{model 1} in low data quality settings.

\section{Conclusion}
We provide the first quantitative analysis demonstrating that when dealing with data from multiple clinical groups, researchers should account for hypothesised group differences in their modelling specifications, and analyse patient \& control groups separately as this provides the most accurate, robust and unbiased recovery of the parameters of interest.

\bibliography{ML4H2020_bibliography}

\begin{thebibliography}{19}
\providecommand{\natexlab}[1]{#1}
\providecommand{\url}[1]{\texttt{#1}}
\expandafter\ifx\csname urlstyle\endcsname\relax
  \providecommand{\doi}[1]{doi: #1}\else
  \providecommand{\doi}{doi: \begingroup \urlstyle{rm}\Url}\fi

\bibitem[Ahn et~al.(2017)Ahn, Haines, and Zhang]{Ahn:2017kp}
Woo-Young Ahn, Nathaniel Haines, and Lei Zhang.
\newblock {Revealing Neurocomputational Mechanisms of Reinforcement Learning
  and Decision-Making With the hBayesDM Package.}
\newblock \emph{Computational psychiatry (Cambridge, Mass.)}, 1:\penalty0
  24--57, October 2017.

\bibitem[Aylward et~al.(2019)Aylward, Valton, Ahn, Bond, Dayan, Roiser, and
  Robinson]{Aylward:2019ka}
Jessica Aylward, Vincent Valton, Woo-Young Ahn, Rebecca~L Bond, Peter Dayan,
  Jonathan~P Roiser, and Oliver~J Robinson.
\newblock {Altered learning under uncertainty in unmedicated mood and anxiety
  disorders}.
\newblock \emph{Nature Human Behaviour}, 3\penalty0 (10):\penalty0 1116--1123,
  2019.

\bibitem[Baker et~al.(2020)Baker, Vilidaite, Lygo, Smith, Flack, Gouws, and
  Andrews]{Baker:2020di}
Daniel~H Baker, Greta Vilidaite, Freya~A Lygo, Anika~K Smith, Tessa~R Flack,
  Andr{\'e}~D Gouws, and Timothy~J Andrews.
\newblock {Power contours: Optimising sample size and precision in experimental
  psychology and human neuroscience.}
\newblock \emph{Psychological methods}, July 2020.

\bibitem[Boehm et~al.(2018)Boehm, Steingroever, and Wagenmakers]{Boehm:2018ez}
Udo Boehm, Helen Steingroever, and Eric-Jan Wagenmakers.
\newblock {Using Bayesian regression to test hypotheses about relationships
  between parameters and covariates in cognitive models.}
\newblock \emph{Behavior research methods}, 50\penalty0 (3):\penalty0
  1248--1269, June 2018.

\bibitem[Brown et~al.(2020)Brown, Chen, Gillan, and Price]{Brown:2020ie}
Vanessa~M Brown, Jiazhou Chen, Claire~M Gillan, and Rebecca~B Price.
\newblock {Improving the Reliability of Computational Analyses: Model-Based
  Planning and Its Relationship With Compulsivity.}
\newblock \emph{Biological psychiatry. Cognitive neuroscience and
  neuroimaging}, 5\penalty0 (6):\penalty0 601--609, June 2020.

\bibitem[Daw(2011)]{Daw:2009p7819}
Nathaniel Daw.
\newblock {Trial-by-trial data analysis using computational models}.
\newblock \emph{Decision Making, Affect, and Learning, Attention and
  Performance XXIII}, 23:\penalty0 1, 2011.

\bibitem[Dayan and Daw(2008)]{Dayan:2008p8403}
Peter Dayan and Nathaniel Daw.
\newblock {Decision theory, reinforcement learning, and the brain.}
\newblock \emph{Cognitive, affective {\&} behavioral neuroscience}, 8\penalty0
  (4):\penalty0 429--453, December 2008.

\bibitem[Gabry et~al.(2017)Gabry, Simpson, Vehtari, Betancourt, and
  Gelman]{Gabry:2017go}
Jonah Gabry, Daniel Simpson, Aki Vehtari, Michael Betancourt, and Andrew
  Gelman.
\newblock {Visualization in Bayesian workflow}.
\newblock \emph{arXiv.org}, \penalty0 (2):\penalty0 389--402, September 2017.

\bibitem[Gelman et~al.(2013)Gelman, Carlin, Stern, Dunson, Vehtari, and
  Rubin]{Gelman:2013una}
Andrew Gelman, John~B Carlin, Hal~S Stern, David~B Dunson, Aki Vehtari, and
  Donald~B Rubin.
\newblock \emph{{Bayesian Data Analysis, Third Edition}}.
\newblock CRC Press, November 2013.

\bibitem[Gershman and Blei(2012)]{Gershman:2012vl}
Samuel~J Gershman and D~M Blei.
\newblock {A tutorial on Bayesian nonparametric models}.
\newblock \emph{Journal of Mathematical Psychology}, 2012.

\bibitem[Huys et~al.(2016)Huys, Maia, and Frank]{Huys:2016jwa}
Quentin J~M Huys, Tiago~V Maia, and Michael~J Frank.
\newblock {Computational psychiatry as a bridge from neuroscience to clinical
  applications.}
\newblock \emph{Nature neuroscience}, 19\penalty0 (3):\penalty0 404--413, March
  2016.

\bibitem[Kruschke(2014)]{Kruschke:2014wl}
John Kruschke.
\newblock \emph{{Doing Bayesian Data Analysis}}.
\newblock A Tutorial with R, JAGS, and Stan. Academic Press, November 2014.

\bibitem[Maia and Frank(2011)]{Anonymous:26lP5Bz9}
Tiago~V Maia and Michael~J Frank.
\newblock {From reinforcement learning models to psychiatric and neurological
  disorders.}
\newblock \emph{Nature neuroscience}, 14\penalty0 (2):\penalty0 154--162,
  February 2011.

\bibitem[Mkrtchian et~al.(2017)Mkrtchian, Aylward, Dayan, Roiser, and
  Robinson]{Mkrtchian:2017hj}
Anahit Mkrtchian, Jessica Aylward, Peter Dayan, Jonathan~P Roiser, and Oliver~J
  Robinson.
\newblock {Modeling Avoidance in Mood and Anxiety Disorders Using Reinforcement
  Learning.}
\newblock \emph{Biological psychiatry}, 82\penalty0 (7):\penalty0 532--539,
  October 2017.

\bibitem[Moutoussis et~al.(2018)Moutoussis, Hopkins, and
  Dolan]{Moutoussis:2018hy}
Michael Moutoussis, Alexandra~K Hopkins, and Raymond~J Dolan.
\newblock {Hypotheses About the Relationship of Cognition With Psychopathology
  Should be Tested by Embedding Them Into Empirical Priors.}
\newblock \emph{Frontiers in psychology}, 9:\penalty0 2504, 2018.

\bibitem[Nord et~al.(2017)Nord, Valton, Wood, and Roiser]{Nord:2017bn}
Camilla~L Nord, Vincent Valton, John Wood, and Jonathan~P Roiser.
\newblock {Power-up: A Reanalysis of 'Power Failure' in Neuroscience Using
  Mixture Modeling.}
\newblock \emph{The Journal of neuroscience : the official journal of the
  Society for Neuroscience}, 37\penalty0 (34):\penalty0 8051--8061, August
  2017.

\bibitem[Seymour et~al.(2012)Seymour, Daw, Roiser, Dayan, and
  Dolan]{Seymour:2012cza}
Ben Seymour, Nathaniel Daw, Jonathan~P Roiser, Peter Dayan, and Raymond~J
  Dolan.
\newblock {Serotonin selectively modulates reward value in human
  decision-making.}
\newblock \emph{The Journal of neuroscience : the official journal of the
  Society for Neuroscience}, 32\penalty0 (17):\penalty0 5833--5842, April 2012.

\bibitem[WHO(2013)]{WHO:2013}
WHO.
\newblock \emph{World\ Health\ Organization: World health report - Mental
  disorders}, 2013.
\newblock URL \url{http://www.who.int/whr/2001/media_centre/press_release/en/}.

\bibitem[Wilson and Collins(2019)]{Wilson:2019kt}
Robert~C Wilson and Anne~Ge Collins.
\newblock {Ten simple rules for the computational modeling of behavioral data.}
\newblock \emph{eLife}, 8:\penalty0 558, November 2019.

\end{thebibliography}

\appendix
\section{Alternative approaches}
Although the issue of proper model specification in computational psychiatry has remained relatively under-researched, there are nonetheless important contributions that should be acknowledged. In particular, it has been previously suggested that the optimal way to investigate symptom-parameter relationships is to build them into the model \citep{Moutoussis:2018hy, Boehm:2018ez}. This approach generally represents group-level parameter values as being offset from a single mean to an extent dependent on a psychiatric covariate (e.g. a continuous psychiatric symptom covariate). However, this could be extended to the dichotomous case by coding the groups as, for example, 0 and 1. We argue that this is largely equivalent to \emph{model 2} which assumes separate priors for different diagnostic groups. However, while this approach may be successful, and avoids shrinkage related issues of under- or over-estimation of effect sizes, it can present various difficulties. For example, when using more than two groups, group identity will need to be dummy coded, which raises the issue of choosing the optimal dummy coding scheme. Second, this approach typically assumes that each group has identical variance. In contrast, our approach allows different groups to have different variances. Third, when multiple covariates are included, as is often the case in psychiatric research where potential confounds must be dealt with, and we wish to model these effects across multiple parameters of interest, model
complexity will increase substantially when this is built into the same model. As a result, parameters will inevitably be less identifiable and may trade off against one another. Separating the estimation of model parameters from the estimation of parameter-symptom relationships avoids this problem (a symptom-agnostic form of analysis).

While the approach of modelling groups separately provides the most accurate parameter estimates, this relies on the availability of known group labels. It may be possible to combine unsupervised clustering methods into the parameter estimation approach to allow the automatic identification of groups within the data \citep{Gershman:2012vl}, obviating the need for diagnostic groups to be manually assigned. In addition to potentially providing more accurate parameter estimates, this could permit the delineation of patient clusters based purely on computational modelling of behaviour.

\begin{figure*}[htbp]
\floatconts
  {fig:models}
  {\caption{Generative model specifications}}
  {\includegraphics[width=1\textwidth]{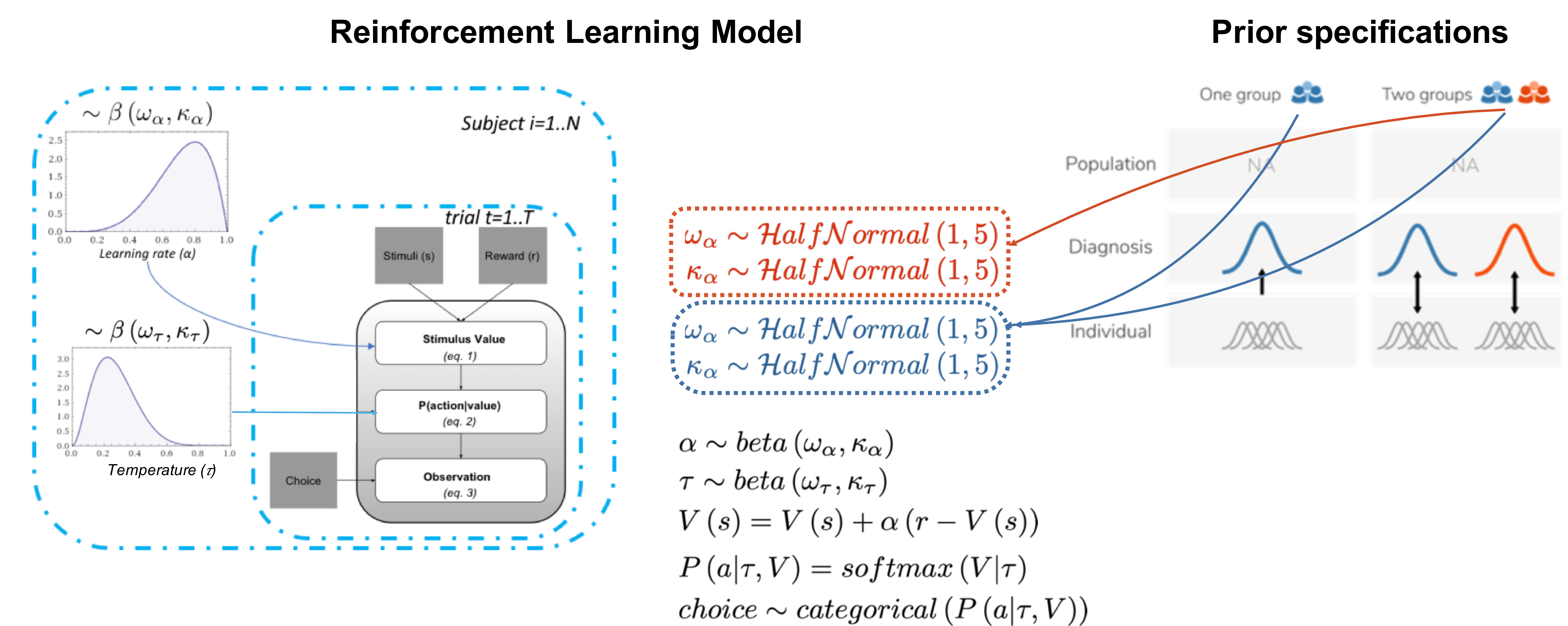}}
\end{figure*}

Finally, another possibility would be to add another layer to the generative model at the population level, which would allow groups to be independent from one another (separate group-level priors) while also forcing the groups to be more similar to one another. This method has proven to be very successful when enough data is available at the different levels of the hierarchy (see Multi-level Regression with Poststratification - \citet{Gelman:2013una}). However, given that the majority of patient studies rely on case-control designs, this would provide insufficient information to estimate the population-level distribution from the data alone (i.e. with just 2 groups), and strong informative hyper-parameters would then be required \citep{Gelman:2013una}. 

\section{RL model specifications}

\emph{Model 1} and \emph{Model 2} are identical, with the exception of group-level priors (see Figure \ref{fig:models}). Under \emph{model 2}, case and controls get their own group-level priors, while \emph{model 1} ignores diagnostic criteria and `pools' all participants together under the same group-level prior. Learning rate `$\alpha$' and temperature `$\tau$' are Beta distributed (all parameters generated were constrained to the range [0,1]). On a given trial, the value of the chosen stimulus is updated as a function of the reward `$r$' received on the current trial, the current expected value for that stimulus `$V(s)$', and the learning rate `$\alpha$'. Probability of choosing an action (bandit arm) follows a Softmax distribution, given stimuli values `$V(s)$' and temperature `$\tau$'.

\section{Using model comparison}

We also attempted to use model comparison as a way to select the model that offered the most accurate fit to the data while being the most parsimonious in terms of model complexity (see Figure \ref{fig:model_comparison}). We used the WAIC (also known as the Widely Applicable Information Criterion  - \citet{Gelman:2013una,Kruschke:2014wl,Ahn:2017kp}) to approximate and compare model evidence. We found that although both models could accurately recover the parameters used to generate the data, model comparison metrics were not sensitive or robust enough to reliably favour one model over the other.

\begin{figure*}[htbp]
\floatconts
  {fig:model_comparison}
  {\caption{Model comparison using WAIC (lowest bar denotes winning model). The winning model is unstable for similarly distributed datasets, yet parameter recovery is equivalent between the two model specifications.}}
  {\includegraphics[width=1\textwidth]{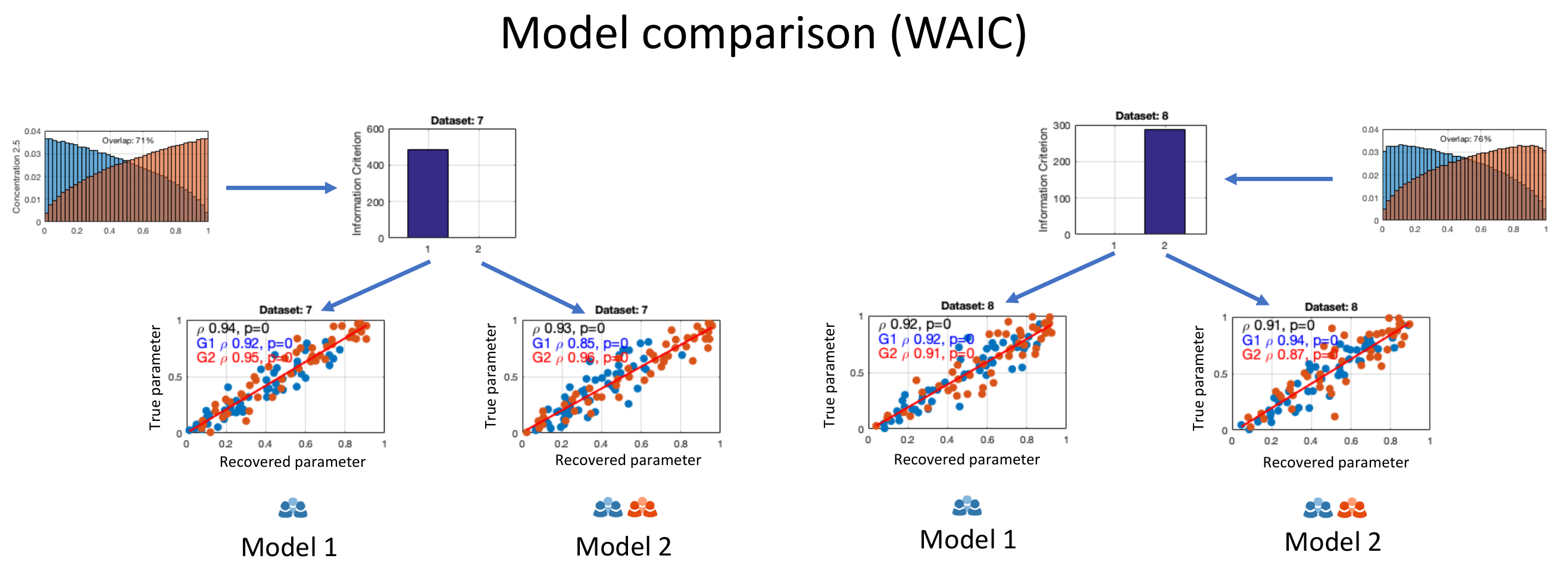}}
\end{figure*}

\end{document}